\documentclass{elsart}

\usepackage{amsmath,amssymb,bm}
\usepackage{natbib,url}

\begin{document}

\begin{frontmatter}

\title{Security Problems with Improper Implementations of Improved FEA-M}
\thanks{This paper has been accepted by The \textit{Journal of Systems \& Software} in May 2005,
and corrected proof has been available online at
\url{http://dx.doi.org/10.1016/j.jss.2006.05.002}.}

\author{Shujun Li\corauthref{corr}} and
\author{Kwok-Tung Lo}
\address{Department of Electronic and Information Engineering, The
Hong Kong Polytechnic University, Hung Hom, Kowloon, Hong Kong SAR,
China}

\corauth[corr]{The corresponding author, personal web site:
\texttt{http://www.hooklee.com}.}

\begin{abstract}
This paper reports security problems with improper implementations
of an improved version of FEA-M (fast encryption algorithm for
multimedia). It is found that an implementation-dependent
differential chosen-plaintext attack or its chosen-ciphertext
counterpart can reveal the secret key of the cryptosystem, if the
involved (pseudo-)random process can be tampered (for example,
through a public time service). The implementation-dependent
differential attack is very efficient in complexity and needs only
$O(n^2)$ chosen plaintext or ciphertext bits. In addition, this
paper also points out a minor security problem with the selection of
the session key. In real implementations of the cryptosystem, these
security problems should be carefully avoided, or the cryptosystem
has to be further enhanced to work under such weak implementations.
\end{abstract}
\begin{keyword}
multimedia encryption \sep FEA-M \sep insecure implementation \sep
differential attack \sep chosen-plaintext attack \sep
chosen-ciphertext attack \sep pseudo-random process
\end{keyword}

\end{frontmatter}

\section{Introduction}

Multimedia data play important roles in today's digital world. In
many multimedia applications, such as pay-TV services, commercial
video conferences and medical imaging systems, fast and secure
encryption methods are required to protect the multimedia contents
against malicious attackers. In recent years, many different
multimedia encryption schemes have been proposed to fulfill such an
increasing demand \citep{Ahl:ImageVideoEncryption:Book2005,
Furht:ImageVideoEncryption:Handbook2004,
ShujunLi:ChaosImageVideoEncryption:Handbook2004}. In
\citep{Yi:FEA-M:IEEETCE2001}, a new fast encryption algorithm for
multimedia (FEA-M) was proposed, which bases the security on the
complexity of solving nonlinear Boolean equations. Later FEA-M was
employed to construct a key agreement protocol by the same authors
in \citep{Yi:FEA-M:IEEETCE2002}. Since then, some attacks of FEA-M
have been reported \citep{Mihaljevic:BreakingFEA-M:IEEECommL2002,
Mihaljevic:BreakingFEA-M:IEEETCE2003,
WuBaoDeng:BreakingFEA-M:LNCS2003,
Youssef:BreakingFEA-M:IEEETCE2003}, most of which can break the key
with a smaller complexity than the simple brute force attack
\citep{Mihaljevic:BreakingFEA-M:IEEECommL2002,
Mihaljevic:BreakingFEA-M:IEEETCE2003,
WuBaoDeng:BreakingFEA-M:LNCS2003}, and one of which can completely
break the whole cryptosystem with only one known and two chosen
plaintext blocks \citep{Youssef:BreakingFEA-M:IEEETCE2003}.

To enhance the security and to avoid some other defects, an improved
version of FEA-M was proposed in
\citep{Mihaljevic:BreakingFEA-M:IEEETCE2003}. This paper reports
some security problems with improper implementations of the
cryptosystem. We point out that the secret key of the cryptosystem
can be revealed by an implementation-dependent differential attack
if the involved (pseudo-)random process can be tampered. One of such
situations is when the pseudo-random process is uniquely controlled
by an external source (such as a public time service), though it
appears that such an implementation would not compromise the
security of the cryptosystem itself. The proposed differential
attack is very efficient, since only two pairs of chosen plaintext
blocks are needed to completely reveal the key. As a result, in a
real implementation of the cryptosystem, it should be ensured that
the embedded pseudo-random process cannot be controlled by illegal
users. Or, the improved FEA-M has to be further enhanced to resist
this implementation-dependent attack. In addition, a minor problem
with the selection of the session key is also discussed in this
paper.

\section{Improved FEA-M}

The original FEA-M \citep{Yi:FEA-M:IEEETCE2001} is a block cipher
with both plaintext and ciphertext feedback. It encrypts the
plaintext in the form of $n\times n$ Boolean matrices, by an
$n\times n$ Boolean key matrix. The elements of the matrices are
either 0 or 1 and all matrix operations are made over $GF(2)$, i.e.,
modulo 2. As a result, the ciphertext is also in the form of
$n\times n$ Boolean matrices.

Previous works have shown that the original FEA-M has the following
defects: 1) the key can be easily broken by an adaptive
chosen-plaintext attack proposed in
\citep{Youssef:BreakingFEA-M:IEEETCE2003}; 2) an efficient
known-plaintext attack can break it with a complexity smaller than
the brute force attack
\citep{Mihaljevic:BreakingFEA-M:IEEECommL2002,
Mihaljevic:BreakingFEA-M:IEEETCE2003,
WuBaoDeng:BreakingFEA-M:LNCS2003}; 3) it is sensitive to packet loss
\citep{Mihaljevic:BreakingFEA-M:IEEETCE2003} and channel errors due
to the use of plaintext feedback.

To overcome the above-mentioned security defects,
\citeauthor{Mihaljevic:BreakingFEA-M:IEEETCE2003} proposed an
improved FEA-M in \citeyear{Mihaljevic:BreakingFEA-M:IEEETCE2003}.
The improved scheme contains two stages: key distribution and
working stage. The first stage generates two $n\times n$ secret
Boolean matrices, a session key $\bm{K}$ and an initial matrix
$\bm{V}$, generally from a master key $\bm{K}_0$, which is also an
$n\times n$ Boolean matrix and known by both the sender and the
receiver. The key distribution protocol is actually the one used in
\citep{Yi:FEA-M:IEEETCE2002} and can be described as follows.
\begin{itemize}
\item
The sender selects $\bm{K}$ and $\bm{V}$ via a (pseudo-)random
process, and computes
\begin{eqnarray}
\bm{K}^* & = &
\bm{K}_0\bm{K}^{-1}\bm{K}_0,\label{equation:distributeK}\\
\bm{V}^* & = & \bm{K}_0\bm{V}\bm{K}_0,\label{equation:distributeV}
\end{eqnarray}
then sends $(\bm{K}^*,\bm{V}^*)$ to the receiver.

\item The receiver recovers $\bm{K}^{-1}$ and $\bm{V}$ by computing
\begin{eqnarray}
\bm{K}^{-1} & = &
\bm{K}_0^{-1}\bm{K}^*\bm{K}_0^{-1},\label{equation:distributeK2}\\
\bm{V} & = &
\bm{K}_0^{-1}\bm{V}^*\bm{K}_0^{-1}.\label{equation:distributeV2}
\end{eqnarray}
\end{itemize}

After the key distribution stage, the sender and the receiver sides
can start the encryption/decryption procedure with the session key
$\bm{K}$ and the initial matrix $\bm{V}$. Denoting the $i$-th
$n\times n$ plain-matrix by $\bm{P}_i$ and the $i$-th $n\times n$
cipher-matrix by $\bm{C}_i$, the encryption procedure is as follows:
\begin{eqnarray}
\bm{C}_i & = &
\bm{K}\left(\bm{P}_i+\bm{K}\bm{V}\bm{K}^i\right)\bm{K}^{n+i}+\bm{K}\bm{V}\bm{K}^i,\label{equation:encryption}
\end{eqnarray}
and the decryption procedure is
\begin{eqnarray}
\bm{P}_i & = &
\bm{K}^{-1}\left(\bm{C}_i+\bm{K}\bm{V}\bm{K}^i\right)\bm{K}^{-(n+i)}+\bm{K}\bm{V}\bm{K}^i.
\end{eqnarray}
The above procedure repeats for each plain/cipher-matrix until the
plaintext/ciphertext exhausts.

\section{Implementation-Dependent Differential Attack}

In this section, we describe an implementation-dependent
differential attack of the improved FEA-M. This attack works under
the conditions that one can tamper the involved (pseudo-)random
process of the improved FEA-M to use the same $\bm{K}$ and $\bm{V}$
in two separate encryption sessions.

Given two plain-matrices, $\bm{P}_i^{(1)}$ and $\bm{P}_i^{(2)}$, and
their corresponding cipher-matrices, $\bm{C}_i^{(1)}$ and
$\bm{C}_i^{(2)}$, we can get Eq.
(\ref{equation:DifferentialAnalysis}).
\begin{eqnarray}
\bm{C}_i^{(1)}+\bm{C}_i^{(2)} & = &
\left(\bm{K}\left(\bm{P}_i^{(1)}+\bm{K}\bm{V}\bm{K}^i\right)\bm{K}^{n+i}+\bm{K}\bm{V}\bm{K}^i\right)\nonumber\\
& & {}
+\left(\bm{K}\left(\bm{P}_i^{(2)}+\bm{K}\bm{V}\bm{K}^i\right)\bm{K}^{n+i}+\bm{K}\bm{V}\bm{K}^i\right)\nonumber\\
& = &
\bm{K}\left(\bm{P}_i^{(1)}+\bm{K}\bm{V}\bm{K}^i\right)\bm{K}^{n+i}
+\bm{K}\left(\bm{P}_i^{(2)}+\bm{K}\bm{V}\bm{K}^i\right)\bm{K}^{n+i}\label{equation:DifferentialAnalysis}\\
& = &
\bm{K}\left(\bm{P}_i^{(1)}+\bm{P}_i^{(2)}\right)\bm{K}^{n+i}\nonumber
\end{eqnarray}
Apparently, Eq. (\ref{equation:DifferentialAnalysis}) means a simple
relation between
$\Delta\bm{C}_i=\bm{C}_i^{(1)}+\bm{C}_i^{(2)}=\bm{C}_i^{(1)}-\bm{C}_i^{(2)}$
and
$\Delta\bm{P}_i=\bm{P}_i^{(1)}+\bm{P}_i^{(2)}=\bm{P}_i^{(1)}-\bm{P}_i^{(2)}$,
i.e., the plaintext and the ciphertext differentials (sums):
\begin{eqnarray}
\Delta\bm{C}_i & = &
\bm{K}\left(\Delta\bm{P}_i\right)\bm{K}^{n+i}.\label{equation:DeltaC=DeltaP}
\end{eqnarray}
As a result, for two consecutive plaintext-matrices, if we choose
$\Delta\bm{P}_{i+1}=\Delta\bm{P}_i$, we can immediately deduce:
\begin{eqnarray}
\Delta\bm{C}_{i+1} & = &
\bm{K}\left(\Delta\bm{P}_{i+1}\right)\bm{K}^{n+i}\nonumber\\
& = & \bm{K}\left(\Delta\bm{P}_i\right)\bm{K}^{n+i}\nonumber\\
& = & \Delta\bm{C}_i\bm{K}.
\end{eqnarray}
Thus, if $\Delta\bm{C}_i$ is invertible, the session key can be
derived easily as follows:
\begin{eqnarray}
\bm{K} & = & \left(\Delta\bm{C}_i\right)^{-1}\Delta\bm{C}_{i+1}.
\end{eqnarray}
To make $\Delta\bm{C}_i$ invertible, one should choose
$\Delta\bm{P}_i$ to be an invertible matrix over $GF(2)$, where note
that $\bm{K}$ is always invertible following the design of the
cryptosystem.

After $\bm{K}$ is broken, one can substitute it into Eq.
(\ref{equation:encryption}) to get a linear equation with $n^2$
unknown variables, i.e., the $n^2$ elements of the initial matrix
$\bm{V}$:
\begin{equation}
\bm{V}\bm{K}^{n+i}+\bm{K}^{-1}\bm{V}=
\bm{K}^{-2}\left(\bm{C}_i-\bm{K}\bm{P}_i\bm{K}^{n+i}\right)\bm{K}^{-i}.\label{equation:solveK-1}
\end{equation}
By solving this linear equation, it is easy to recover $\bm{V}$.
Actually, we can further reduce the linear equation to directly
deduce $\bm{V}$. Choosing two continuous plaintext matrices
$\bm{P}_i$, $\bm{P}_j$ and adding the two linear systems, one has
\begin{eqnarray}
\bm{V}\bm{K}^{n+i}\left(\bm{I}+\bm{K}^{j-i}\right) & = &
\bm{K}^{-2}\left(\bm{C}_i-\bm{K}\bm{P}_i\bm{K}^{n+i}\right)\bm{K}^{-i}\nonumber\\
& & {}+
\bm{K}^{-2}\left(\bm{C}_j-\bm{K}\bm{P}_j\bm{K}^{n+j}\right)\bm{K}^{-j}.\label{equation:solveK-2}
\end{eqnarray}
When $\bm{I}+\bm{K}^{j-i}$ is invertible, $\bm{V}$ can be
immediately solved by multiplying the right side by
$\left(\bm{I}+\bm{K}^{j-i}\right)^{-1}\bm{K}^{-(n+i)}$ at the end.
Note that $\bm{K}^{n+i}+\bm{K}^{n+j}$ may never be invertible over
$GF(2)$ (for example, when $\bm{K}=\bm{I}$), though the probability
is relatively small when $n$ is relatively high. Once such an event
occurs, one can turn to solve Eq. (\ref{equation:solveK-1}). If
$\bm{V}$ can still not be solved from Eq. (\ref{equation:solveK-1}),
one has to carry out the attack with some other different values of
$\bm{K}$ until $\bm{V}$ can be uniquely solved.

Once $\bm{K}$ and $\bm{V}$ are both known, one can use the method
proposed in Sec. III of \citep{Youssef:BreakingFEA-M:IEEETCE2003} to
recover the master key $\bm{K}_0$.

To carry out a successful attack, in most cases, the attacker only
needs to choose two plaintexts with four chosen plaintext matrices,
$\bm{P}_i^{(1)}$, $\bm{P}_{i+1}^{(1)}$, $\bm{P}_i^{(2)}$ and
$\bm{P}_{i+1}^{(2)}$, which satisfy
$\bm{P}_{i+1}^{(1)}-\bm{P}_{i+1}^{(2)}=\bm{P}_i^{(1)}-\bm{P}_i^{(2)}=\Delta\bm{P}$
and $\Delta\bm{P}$ is an invertible matrix. Considering each matrix
is a $n\times n$ Boolean matrix, $4n^2$ chosen plain-bits are
required in total. When $n=64$, as suggested in
\citep{Yi:FEA-M:IEEETCE2001, Yi:FEA-M:IEEETCE2002}, only 2048
plain-bytes are needed. In addition, the complexity of the proposed
attack is very small, actually it is of the same order as the one
proposed in \citep{Youssef:BreakingFEA-M:IEEETCE2003}. In the case
that $\bm{V}$ can not be solved with four chosen plaintext matrices,
more plaintext matrices have to be chosen, but the number of chosen
plaintext bits is still of the same order -- $O(n^2)$.

Next, let us see in which improper implementations an attacker can
manage to tamper the involved (pseudo-)random process to activate
the above differential attack. Apparently, the above attack requires
two encryption sessions with the same session key $\bm{K}$ and the
same initial matrix $\bm{V}$, one for encrypting the first plaintext
$\left\{\cdots,\bm{P}_i^{(1)},\bm{P}_{i+1}^{(1)}\right\}$ and the
other for encrypting the second plaintext
$\left\{\cdots,\bm{P}_i^{(2)},\bm{P}_{i+1}^{(2)}\right\}$. However,
in each encryption session, $\bm{K}$ and $\bm{V}$ have to be reset
at the sender side via a (pseudo-)random process and distributed to
the receiver side via the key distribution protocol. As a result,
generally two different sessions use different $\bm{K}$ and
$\bm{V}$. However, in real world the encryption scheme may be
improperly implemented such that the attacker can tamper the
(pseudo-)random process. As a typical example, let us assume that
the process is uniquely determined by the system clock\footnote{In
\citep{Yi:FEA-M:IEEETCE2001, Yi:FEA-M:IEEETCE2002,
Mihaljevic:BreakingFEA-M:IEEETCE2003}, it is not mentioned how to
realize the random process. One of the simplest (though maybe less
frequently-used) method to realize a pseudo-random process is to
initialize the seed of the pseudo-random number generator using the
current time stamp. A list of some other more complicated ways can
be found in Section ``The Collection of Data Used to Create a Seed
for Random Number" of \citep{RSAENH2005}.}. In chosen-plaintext
attacks, the attacker has a temporary access to the encryption
machine, so he can intentionally alter the system clock to control
the (pseudo-)random process before running each session to get the
same $\bm{K}$ and $\bm{V}$ for two separate sessions. In addition,
if the improved FEA-M is implemented in such an insecure way that
the second stage can restart without running the key distribution
stage, the attack becomes straightforward.

At last, it deserves mentioned that the above differential
chosen-plaintext attack can be easily to generalize to a
differential chosen-ciphertext attack, provided that the
(pseudo-)random process at the decryption machine can be tampered.
Rewrite Eq. (\ref{equation:DeltaC=DeltaP}) into the following form:
\begin{eqnarray}
\Delta\bm{P}_i & = &
\bm{K}^{-1}\left(\Delta\bm{C}_i\right)\bm{K}^{-(n+i)}.
\end{eqnarray}
Then, by choosing $\Delta\bm{C}_{i+1}=\Delta\bm{C}_i$, one has
\begin{eqnarray}
\Delta\bm{P}_{i+1} & = &
\bm{K}^{-1}\left(\Delta\bm{C}_{i+1}\right)\bm{K}^{-(n+i+1)}\nonumber\\
& = & \bm{K}^{-1}\left(\Delta\bm{C}_i\right)\bm{K}^{-(n+i)-1}\nonumber\\
& = & \Delta\bm{P}_i\bm{K}^{-1}.
\end{eqnarray}
Other steps are identical with the above differential
chosen-plaintext attack.

\section{A Minor Problem with Selection of Session Key}

It is noticed that $\bm{K}$ cannot be selected at random from all
invertible matrices over $GF(2)$. Since all $n\times n$ invertible
matrices form a general linear group $GL(n,2)$, whose order is
$O=\prod_{i=0}^{n-1}(2^n-2^i)$ \citep{GLG}. So, denoting the order
of $\bm{K}$ over $GL(n,2)$ by $o(\bm{K})$, it is true that
$o(\bm{K})\mid O$, i.e., $\bm{K}^{o(\bm{K})}=\bm{I}$, where $\bm{I}$
is the identity Boolean matrix \citep{Gilbert:Algebra2005}. It is
obvious that $o(\bm{K})$ actually corresponds to the periodicity of
the encryption/decryption function with respect to the
plaintext/ciphertext index $i$. Generally speaking, the periodicity
should not be too small to maintain an acceptable security level. As
an extreme example, when $\bm{K}=\bm{I}$, $o(\bm{K})=1$ and the
encryption procedure becomes $\bm{C}_i=\bm{P}_i$ (the cipher
vanishes). Thus, $\bm{K}$ should be selected randomly from all
invertible Boolean matrices with sufficiently large orders, which
means a significant reduction of the session key space.

\section{Conclusions}

This paper reports an implementation-dependent differential attack
of an improved fast encryption algorithm for multimedia (FEA-M)
proposed in \citep{Mihaljevic:BreakingFEA-M:IEEETCE2003}. The attack
works under the condition where the involved (pseudo-)random process
can be tampered by the attacker. In this case, the attack can reveal
the key with four or more chosen plaintext/ciphertext matrices,
i.e., $4n^2$ chosen plain/ciphertext bits, in two or more separate
encryption sessions. The result shows that a secure cryptosystem may
become totally insecure with seemingly-harmless implementation
details in real world \citep{Schneier:Secrets&Lies2000}. In
addition, a minor problem with the selection of the session key is
also discussed in this paper.

\section{Acknowledgements}

This research was supported by The Hong Kong Polytechnic
University's Postdoctoral Fellowships Scheme under grant no. G-YX63.
The authors thank the anonymous reviewers for their valuable
comments to enhance the quality of this paper.

\bibliographystyle{elsart-harv}
\bibliography{FEA-M}

\end{document}